\documentclass[%
 reprint,
 superscriptaddress,
 amsmath,amssymb,
 aps,
]{revtex4-1}

\usepackage{graphicx}
\usepackage{dcolumn}
\usepackage{bm}

\usepackage[pdf]{pstricks}
\usepackage{amsmath}
\usepackage{amsfonts}
\usepackage{amssymb}
\usepackage{graphicx}
\usepackage{float}
\usepackage[caption = false]{subfig}
\usepackage{braket}
\usepackage{multirow}
\usepackage{amsmath}
\newcommand{\change}[1]{\textcolor{black}{#1}}
\newcommand{\changeSecondSubmission}[1]{\textcolor{black}{#1}}

\begin{document}

\preprint{APS/123-QED}

\title{Stimulated resonant inelastic x-ray scattering with chirped, broadband pulses}


\author{Maximilian Hollstein}
\affiliation{Max Planck Institute for the Structure and Dynamics of Matter, Luruper Chaussee 149, 22761 Hamburg, Germany}
 \affiliation{Center for Free-Electron Laser Science, DESY, Notkestrasse 85, 22607 Hamburg, Germany
}
\author{Nina Rohringer}%
 \affiliation{Max Planck Institute for the Structure and Dynamics of Matter, Luruper Chaussee 149, 22761 Hamburg, Germany}
\affiliation{Center for Free-Electron Laser Science, DESY, Notkestrasse 85, 22607 Hamburg, Germany
}
\affiliation{Department of Physics, University of Hamburg, Jungiusstrasse 9, 20355 Hamburg, Germany
}




\date{\today}

\begin{abstract}
We present an approach for initiating and tracing ultra-fast electron dynamics in core-excited atoms, molecules and solids. The approach is based on stimulated resonant inelastic x-ray scattering induced by a single, chirped, broadband XUV/x-ray pulse. A first interaction with this pulse prepares a core-excited state wave packet by resonant core-excitation. A second interaction with the pulse at a later time induces the transition to valence-excited states which is associated with stimulated emission. The preparation of the core-excited wave packet and the transition from the core-excited states to  the valence-excited states occur at distinct chirp-dependent times. As a consequence, the stimulated emission carries information about the time evolution of the core-excited state wave packet.
\end{abstract}

\maketitle

\section{Introduction}
The availability of femtosecond and attosecond  pulses at free electron lasers (FEL) \cite{Ackermann2007,RevModPhys.88.015007,PhysRevLett.119.154801} and at table-top sources  based on the high-harmonic generation process \cite{Hentschel2001} makes the observation of ultra-fast processes on attosecond and few-femtosecond timescales possible.
This is paving the way towards the investigation of fundamental aspects associated with the purely electronic response of molecular systems to prompt perturbations. It is anticipated that experiments using these novel light sources will provide insight into the first steps of chemical reactions that are governed by purely electronic motion \cite{CEDERBAUM1999205,0953-4075-47-12-124002,PhysRevLett.94.033901,Remacle2006}. 

So far, experiments that have aimed at ultra-fast electron dynamics and nuclear dynamics were based on \change{high harmonic spectroscopy \cite{Smirnova2009, Woerner_2011_0009_4293_299,PhysRevLett.104.233904,Haessler2010,Kraus790}, XUV-induced fragmentation \cite{PhysRevLett.110.053003,doi:10.1021/acs.jpclett.5b01205,Rudenko2017,PhysRevX.6.021035,0953-4075-51-3-032003} combined with ion charge and momentum spectroscopy, XUV-pump-XUV-probe schemes \cite{PhysRevLett.113.073001} as well as on combinations of a NIR pulse and an XUV pulse that arrive at the sample under investigation with a variable time-delay \cite{Drescher2002,Ott2014,Calegari336,Goulielmakis2010,WIRTH2013149,Pertot264,doi:10.1063/1.4898375}}.   In this context, broadband XUV pulses have been used to initiate electron dynamics which was subsequently probed using a time-delayed NIR pulse by means of streaking \cite{Drescher2002}, modification of the dipole response with the NIR \cite{Ott2014,Ott716} and  mass spectrometry after NIR-induced fragmentation \cite{Calegari336}. Also other schemes, where the pulses switched role, have been considered to initiate and trace electron dynamics. For instance, NIR pulses have been employed to initiate electron and nuclear dynamics which were subsequently probed by transient absorption spectroscopy using a time-delayed XUV pulse \cite{Goulielmakis2010,WIRTH2013149,Pertot264,doi:10.1063/1.4898375} or mass spectrometry after XUV-induced dissociation \cite{Erk288}.
\changeSecondSubmission{Here, we suggest to employ chirped x-ray/XUV pulses to inititate and probe electron dynamics. Chirped visible laser pulses have been previously explored to initiate and control nuclear dynamics \cite{PhysRevLett.74.3360,AMSTRUP199487,legare}. The modulation of phases and spectral components in photo excitation pulses has been demonstrated to provide optical control of the primary step of photoisomerization \cite{Prokhorenko1257,Joffre453} and theoretical investigations have considered various types of spectroscopies that are suitable to trace non-adiabatic dynamics at conical intersections \cite{doi:10.1021/acs.chemrev.7b00081}. }

 To transfer these kind of studies to the sub-femtosecond regime, attosecond-pump-attosecond-probe experiments are anticipated to be of importance \cite{Leone2014}. Here, pump-probe spectroscopy and multidimensional spectroscopy \cite{Mukamel2013} with attosecond XUV/x-ray pulses are particularly promising since it allows one to trace dynamics with both high spectral  and high temporal resolution \cite{Goulielmakis2010,WIRTH2013149,PhysRevA.83.033405}. Moreover, the element-specificity of core-excitations can give rise to spatial sensitivity in hetero-atomic systems. This feature is particularly advantageous in the context of tracing ultrafast charge transfer processes in molecular systems \cite{PhysRevLett.89.043001,PhysRevA.76.012504,doi:10.1063/1.3557057,PhysRevA.88.013419,PhysRevA.90.023414}. In contrast to indirect and ambiguous probing of electron dynamics by, for instance, photo-induced fragmentation  which involves nuclear motion \cite{Calegari336}, spectroscopy allows for the reconstruction of the time-dependent electronic state \cite{Goulielmakis2010} or observables that are directly related to the electron dynamics \cite{PhysRevA.88.013419,PhysRevA.90.023414,PhysRevA.95.053411}. 

To our best knowledge, attosecond-pump-attosecond-probe experiments tracing attosecond electron dynamics have not yet been conducted. This appears to be predominantly due to the low photon flux of the attosecond pulses that have been available. This has prevented their use to induce  nonlinear processes.  
Recently, however, the production of intense attosecond pulse with pulse energies in the range of nJ to $\mu$J that are suitable for this purpose has been demonstrated at HHG \cite{Ferrari2010,Takahashi2013,Sansone2011,PhysRevLett.120.093002} sources as well as at FELs \cite{PhysRevLett.119.154801}. The implementation of attosecond-pump-attosecond-probe experiments can therefore be anticipated for the near future.

Still, the time-resolution achievable in pump-probe experiments based on these intense attosecond pulses is suboptimal due to properties of the pulses originating from their specific creation process. For instance, pulses created by the HHG process are intrinsically chirped \cite{PhysRevLett.102.093002}. Thus, these pulses are not as short as their often very large bandwidth would allow for. This affects the time resolution achievable in experiments using these pulses. Also the time-resolution achievable in pump-probe experiments at FELs is limited by the generation process. In particular at FELs that are based on the self amplified spontaneous emission (SASE) process, the arrival time of the pulses strongly fluctuates \cite{RevModPhys.88.015007}. This complicates the synchronization of two SASE pulses and thus limits the time resolution achievable in pump-probe experiments using them. 

In this work, we present an alternative spectroscopic technique applicable to tracing ultra-fast dynamics upon core-excitation and which improves for this particular purpose the time resolution achievable with the ultra-short pulses available. The technique presented is based on single, linearly chirped, broadband pulses which are used to obtain dynamical information of atoms or molecules with sub-pulse duration time resolution. The approach can therefore be implemented by pulses created by the HHG process which are intrinsically linearly chirped. 
Another interesting feature of the approach represents the fact that it is based on single, isolated pulses. This means in particular that a synchronization of two pulses, as in conventional pump-probe experiments, is not necessary. Therefore, the approach is  insensitive to the arrival time jitter that is present, for instance, at SASE FELs. In this context, it should be noted that FEL pulses can be chirped \cite{RevModPhys.88.015007}. Hence, the approach presented might be used to push the achievable time resolution and the dynamics that can be investigated at table-top sources as well as at FEL towards shorter timescales.

The technique presented is based on the stimulated resonant inelastic XUV/x-ray scattering process (SRIXS) induced by a broadband, linearly chirped XUV/x-ray pulse. SRIXS is an inelastic two-photon process in which a photon with frequency $\omega_1$ is absorbed causing the promotion of a core electron to an unoccupied orbital. A second interaction with the light field causes subsequently the refilling of the core-hole  by a  valence electron, which, in the following, will be  referred to as 'dump' transition. In the course of the process, a photon with a second frequency $\omega_2 \neq \omega_1$ is emitted.

\begin{figure}
\centering
\subfloat[][]{\includegraphics[width=1\linewidth]{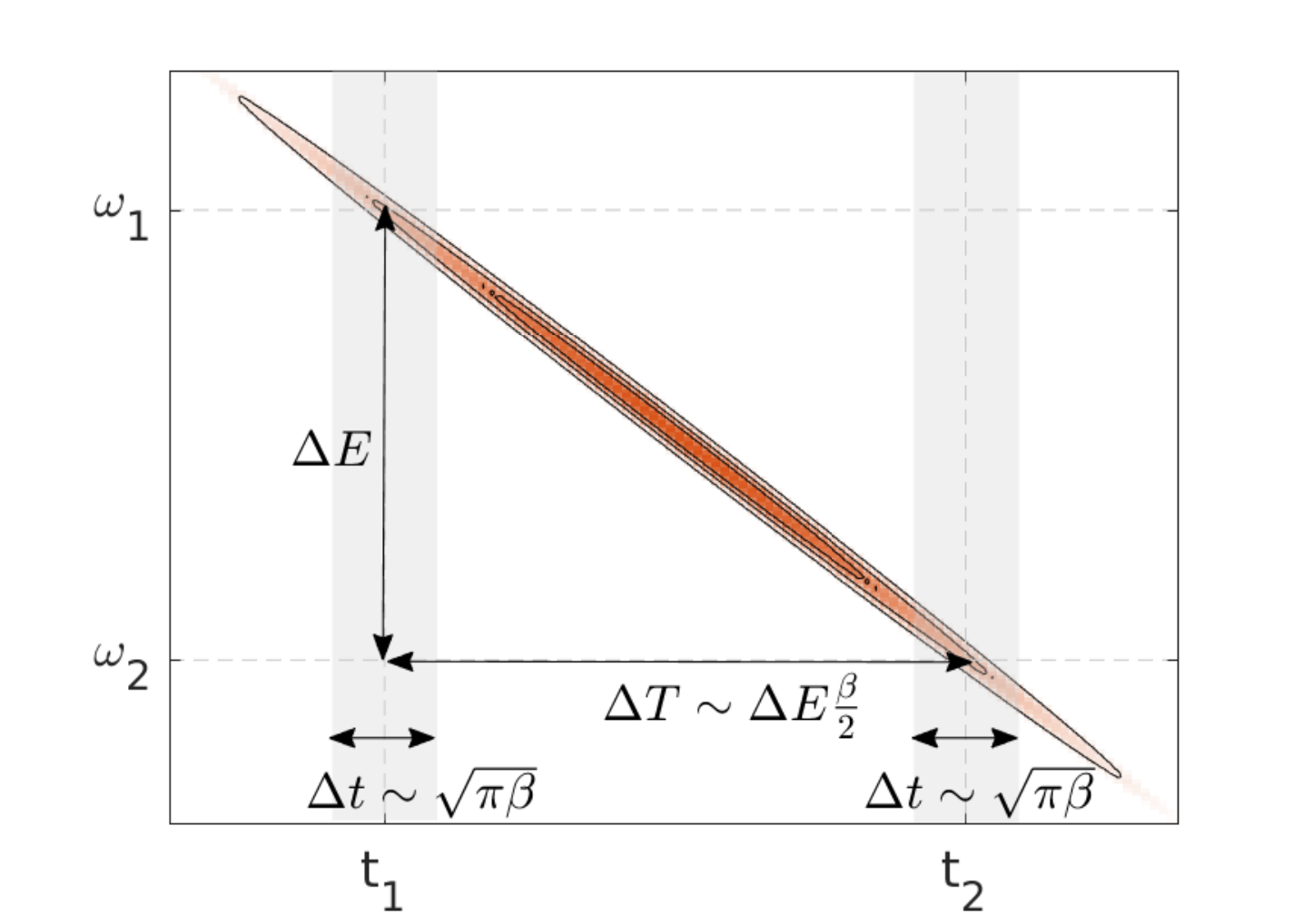}}
\qquad
\subfloat[][]{\includegraphics[width=0.45\textwidth]{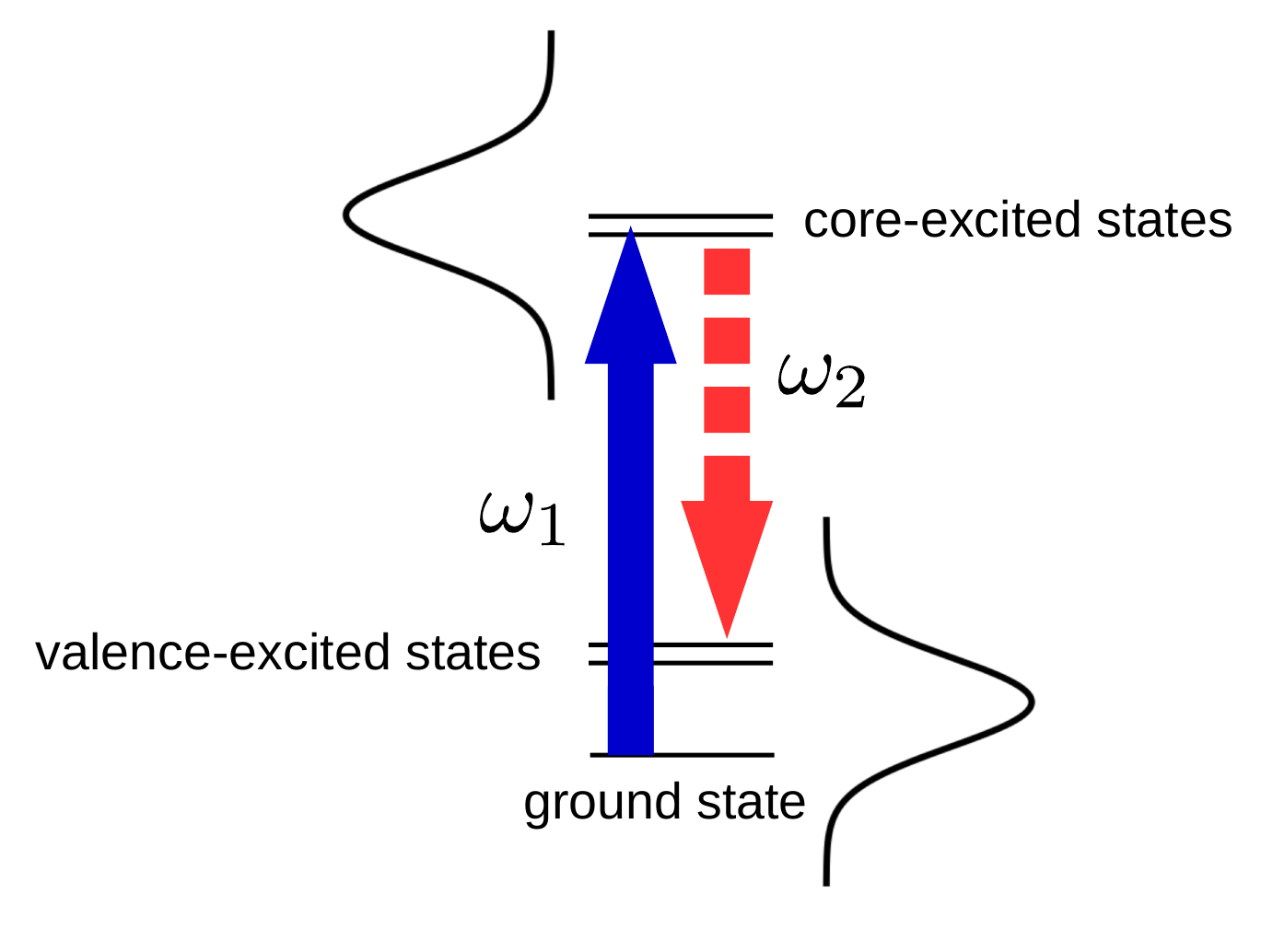}}
\caption{\label{fig:wignerdistribution}(a) Wigner distribution of a chirped electric field of the form given in Eq. \ref{eq:chirped_pulse}. (b) Stimulated resonant inelastic x-ray scattering. The field induces transitions with transition energies $\omega_1$ associated with core-excitations (blue arrow with solid line in (b)) and $\omega_2$ associated with dump transitions (red arrow with dashed line in (b)) at different times.  }

\end{figure}

The key idea of the approach presented is based on the fact that in a chirped pulse, different frequencies arrive at different times at the system exposed to the field. Using broadband pulses, energetically distinct classes of transitions can be induced at distinct times. As it has been recently discussed in the context of two-photon ionization \cite{PhysRevA.95.043424}, this can be used to realize pump-probe experiments with a single pulse. In the context of SRIXS, as it is shown in this work and sketched in Fig. \ref{fig:wignerdistribution}, it can be used to prepare core-excited state wave packets and to probe them  by the dump transition  after a chirp-dependent time-delay.  As it is demonstrated in the following, the dynamics associated with the wave packet of core-excited states are then imprinted in the photo emission that accompanies the refilling of the core hole.  In this way, ultra-fast electronic processes following core-excitation such as charge migration in  molecules  can be evidenced and investigated with sub-pulse-duration time resolution.

This paper is structured as follows: first, we develop, based on time-dependent perturbation theory, a theory of SRIXS induced by linearly chirped, broadband pulses. This approximate theory shows the potential of the approach to probe ultra-fast electron dynamics upon core-excitation.  Second, we validate the approximate theory by means of numerical simulations concerning argon atoms using the Maxwell-Bloch approach described in Ref. \cite{PhysRevA.90.063828}. Thereby, we demonstrate the applicability of the approach to initiate and to monitor few-fs  electron dynamics  upon resonant core-excitation in atoms.

\section{Theory of SRIXS with chirped, broadband pulses}
In the following, we develop a theory of  SRIXS induced by chirped broadband x-ray/XUV pulses. For this purpose, we first consider the dynamics that are induced by a chirped field in few level systems representing atoms or molecules. On this basis, we determine the time-dependent polarization that enters Maxwell's equations which we subsequently solve approximately to obtain the spectrum of the transmitted light.  If not stated otherwise, atomic units will be used.
\subsection{Field-induced dynamics}
In the following, we focus our considerations on linearly chirped pulses with electric fields that have the form:\begin{equation}\label{eq:chirped_pulse}
E(t)=\frac{1}{2}\big[\mathcal{E}(t)e^{i(\omega_0-\frac{t}{\beta})t}+c.c.\big].
\end{equation} Here, $\omega_0$ represents the carrier frequency,  $\mathcal{E}(t)$ denotes an envelope which varies slowly in time with respect to the phase factor  $e^{i\omega_{0}t}$ and $\beta > 0$ determines the magnitude of the chirp. 
In the following, we consider a few level system that may represent an atom or a molecule whose state is described by the wave function $|\psi(\tau)\rangle$:
\begin{equation}\label{eq:time_dependent_wave_function}|\psi(t)\rangle = \alpha_{0}(t)|0\rangle + \sum_{I}\alpha_I(t)|I\rangle + \sum_{F}\alpha_F(t)|F\rangle.  \end{equation} Here, $|0\rangle$ denotes the ground state, $|I\rangle$ represent core-excited states that are dipole-coupled to the ground state, $|F\rangle$ are valence-excited states that are dipole-coupled to the core-excited states and  $\alpha_{0}(t),\alpha_{I}(t),\alpha_{F}(t)$ represent in the subsequent considerations state amplitudes in the interaction picture.
\subsubsection{Transition amplitudes of core-excited states $\alpha_I$} To determine the amplitudes $\alpha_I$ approximately, we consider the transition amplitude in first-order time-dependent perturbation theory:\begin{equation}\label{eq:FOTA}
\alpha_I(t)\sim\frac{1}{2i}d_{I,0}\int_{-\infty}^{t}dt'\big[\mathcal{E}(t')e^{i (\omega_0-\frac{t'}{\beta})t'}+c.c.\big]e^{i(E_I - E_0)t'}
\end{equation}
where $d_{I,0}$ represent dipole transition matrix elements between the core excited state $|I\rangle$  to the ground state and $E_I$, $E_0$ represent the energies of $|I\rangle$ and the ground state, respectively. 
Noting that the integral given in Eq. \ref{eq:FOTA} has only significant values when $t$ is on the order or larger than $t_I = (\omega_0 - E_I + E_0) \frac{\beta}{2}$, one finds that, as one intuitively expects, the interaction with the chirped pulse induces the core-excitations in the vicinity of times at which the instantaneous carrier frequency \change{$\omega_{inst}(t) = \omega_0 -\frac{2t}{\beta}$} is resonant with the respective transitions $(\omega_{inst}(t_I) = E_I - E_0)$. For times $t$ much larger than $\sim t_I + \sqrt{\beta}$, the amplitudes  can be found \changeSecondSubmission{(for details, see the Appendix 1)} to be approximatly: \begin{equation}\label{eq:tdfoc} \alpha_I(t\gg t_I + \sqrt{\beta})\sim -\frac{1}{2i}\sqrt{\pi\beta} d_{I,0}\mathcal{\overline{E}}(t_I)e^{-i\frac{t_I^{2}}{\beta}}e^{i\frac{\pi}{4}}.\end{equation} Note that this expression coincides with the transition amplitude associated with the interaction of the system with a resonant field with a constant amplitude $\mathcal{\overline{E}}(t_I)e^{-i\frac{t_I^{2}}{\beta}}e^{i\frac{\pi}{4}}$ during a time interval $\Delta t$ with length $\sqrt{\pi\beta}$.  This can intuitively  be understood  when noting that the energy uncertainty associated with this time interval via the time-energy uncertainty relation is on the order of the change of the instantaneous carrier frequency during this very same time interval.
\change{
\begin{equation}
\Delta\omega :=|\omega_{inst}(t) -\omega_{inst}(t+\sqrt{\pi\beta})|= 2\sqrt{\frac{\pi}{\beta}} = \frac{2\pi}{\Delta t}
\end{equation}}

That is, the excitations effectively occur during  time intervals during which the instantaneous carrier frequency is resonant with the respective transition energy up to the energy uncertainty. 

\subsubsection{Transition amplitudes of valence-excited states $\alpha_F$}
Under the same assumptions, analogous considerations based on second-order time-dependent perturbation theory \changeSecondSubmission{(for details, see the Appendix 2)} concerning the dump transitions show that these are induced by the field in the vicinity of $t_{F,I} =  (E_F - E_I + \omega_0)\frac{\beta}{2}  $, i.e., at times when the instantaneous carrier frequency is resonant with the dump-transitions.  For $t_{F,I} - t_{I} > \sqrt{\pi\beta}$, the core-excitation and the dump transition are predominantly induced during non-overlapping time intervals so that in this situation, the transition amplitudes can be approximated in terms of products of the amplitudes of the core-excited states (Eq. \ref{eq:tdfoc}) and first order transition amplitudes from core-excited states to the valence excited states. This yields:

\change{
\begin{multline}\label{approx:transition_amplitude_final_states}
\alpha_F(t \gg t_{F,I}+ \frac{\sqrt{\pi\beta}}{2})\sim\\-\frac{1}{2i}e^{i\frac{t_{F,I}^{2}}{\beta} }\sqrt{\pi\beta}e^{-i\frac{\pi}{4}}\mathcal{E}(t_{F,I})\sum_{I}\alpha_I(t\gg t_1+\sqrt{\beta})d_{F,I}
\end{multline}}

Here, $d_{F,I}$ are dipole matrix elements between the valence excited state $|F\rangle$ and the core excited state $|I\rangle$.
\subsubsection{Impulsive limit}
On the basis of the previous considerations, one can determine the limit in which the interaction with a chirped pulse may represent a pump-probe experiment with a well defined time delay.
According to Eq. \ref{eq:tdfoc}, the times $t_I$ may be replaced by an averaged excitation time $t_I \rightarrow t_1 = \langle t_I\rangle_I$ (here, $\langle...\rangle_I$ represents averaging over all core-excited states involved) if the temporal range spanned by the times $t_I$ is small in comparison to the dynamical timescales being probed, small in comparison to $\sqrt{\beta}$ as well as small in comparison to the characteristic timescale on which the slowly varying envelope $\mathcal{E}$ changes. Under the analogous assumptions if the level spacings between the valence excited states are much smaller than the level spacing between ground state and valence excited states, the times $t_{F,I}$ may be replaced by an average deexcitation time $t_2 \rightarrow \langle t_{F,I}\rangle_{F,I}$.
In this limit, the interaction with the field is effectively impulsive. That is, the preparation of the core-excited state wave packet as well as its probe by the dump transitions occur effectively promptly with respect to the relevant timescales and a well-defined  pump-probe time-delay $t_2 - t_1$ exists.

%
%
%

\subsection{Spectrum of the transmitted light}
We now turn to the spectrum of the transmitted field. For this, we approximately solve Maxwell's equations for the situation considered. In the slowly varying envelope approximation, they reduce to an equation for the spectral envelope of the field $\tilde{\mathcal{E}}$ (see for instance Ref. \cite{PhysRevA.83.033405})
 \begin{equation}\label{eq:mwe_SVEA}
 \frac{\partial\tilde{\mathcal{E}}(z,\omega)}{\partial z} = 2\pi i\frac{\omega}{c}\tilde{\mathcal{P}}(z,\omega)
 \end{equation} where $\tilde{\mathcal{P}}(z,\omega)$ denotes the Fourier transform of the polarization $\mathcal{P}(z,\tau) = \rho\langle\psi(z,t)|D|\psi(z,t)\rangle$, $c$ represents the velocity of light, $\rho$ is the atomic number density, $D$ is the dipole operator and  $\psi(z,\tau)$ is the electronic wave function of an atom at position $z$.

\change{In the impulsive limit, under the assumptions that the field keeps the form given in Eq. \ref{eq:chirped_pulse} and the excitation probability of core-excited states remains mostly propagation distance independent, Eq. \ref{eq:mwe_SVEA} can be integrated analytically \changeSecondSubmission{(for details, see the Appendix 3)} yielding the spectral envelope of the field after the transmission through the sample $\mathcal{E}(L,\omega)$ at frequencies in the vicinity of the dump-transition energies:
\begin{equation} |\tilde{\mathcal{E}}(L,\omega)|^2 \sim |\tilde{\mathcal{E}}(0,\omega)|^2 e^{\sigma(\omega)\rho L}. \end{equation}
Here, $\sigma$ represents the emission cross section which is given by:}

\changeSecondSubmission{
\begin{multline}\label{eq:difference_spectrum}\sigma =  -\frac{\pi^2\omega\beta|\mathcal{E}(t_1)|^2}{c}\text{Im}\big[\sum_{I,I',F} \\ d_{0,I'}d_{I,0}
d_{I',F}d_{F,I}e^{-i(E_I-E_{I'})(t_{2} - t_{1})}\\\frac{1}{E_F-E_I + \omega +\frac{i\Gamma_I}{2}} e^{-\frac{\Gamma_I (t_{2} - t_1)}{2}}e^{-\frac{\Gamma_{I'} (t_{2} - t_{1})}{2}}\big];
\end{multline} 
}\changeSecondSubmission{
$L$ denotes the propagation distance and $\Gamma_{I}$ represent the decay rates of the core-excited states $|I\rangle$.
It is worth noticing that the appearance of the phase factors $e^{-i(E_{I}-E_{I'})(t_{2}-t_{1})}$ in Eq. \ref{eq:difference_spectrum}}, which are related to the time evolution of the core-excited state wave packet, shows the potential of the approach to initiate and to probe ultra-fast electron dynamics upon resonant core excitation.

\subsection{Requirements}
In the following, we discuss the requirements of the approach concerning \textbf{(I)} the systems to which it can be applied to and \textbf{(II)} concerning the pulse parameters that are needed to realize the approach.

\begin{itemize}\item[\textbf{(I)}]
For the above considerations to be valid, the level spacing between the core-excited states and between the valence excited states, respectively, has to be much smaller than the level spacing between ground state and valence excited states. Typically, the level spacing between core-excited states as well as between valence-excited states, is on the order of $\sim 1 \,$eV whereas the spacing between ground state and valence excited states is usually on the order of $\sim 10\,$ eV to several tens of eV.
As indicated by the numerical simulations concerning argon atoms discussed below, this can be sufficient.
\item[\textbf{(II)}]Concerning the pulse parameters, the above considerations indicate that a pump-probe experiment associated with a finite time delay, can be implemented using single, linearly chirped pulses inducing the SRIXS process if:
\begin{itemize}
\item[\textbf{(a)}] The bandwidth of the pulse exceeds the energy gap $\Delta E$ between ground state and valence-excited states.
\item[\textbf{(b)}] $\Delta E\frac{\beta}{2}$ (the effective time-delay) is larger than the time resolution $\sqrt{\pi\beta}$ achievable representing  a constraint to the chirp parameter: $\beta > \frac{4\pi}{\Delta E^2}$.
\item[\textbf{(c)}] The dynamical timescales being probed exceed the achievable time-resolution $\sqrt{\pi\beta}$. 

\end{itemize}

Noting that the level spacing between ground state and valence-excited states in atoms, molecules and solids is typically on the order of one to a few tens of eV.  This energy spacing is covered by the bandwidth of broadband pulses created by the high-harmonic generation process (HHG) which is typically in the range of 10 eV to even more than 100 eV \cite{Chini2014}. Moreover, the  pulses produced by the HHG process may also fulfill the constraint $\beta > \frac{4\pi}{\Delta E^2}$ (II)(b). For instance, for $\Delta E\sim 1\,$ a.u. (which is for instance often realized for valence excitations involving inner-valence electrons), $\beta$ has to be larger than 12 $\frac{as}{eV}$. The pulses originating from HHG are intrinsically chirped. Depending on the driver wavelength and the driver intensity \cite{Mairesse1540,PhysRevLett.102.093002}, it has been shown that their chirp $\beta$ can be varied from 8$\frac{as}{eV}$ to 41$\frac{as}{eV}$ \cite{PhysRevLett.102.093002}.  Hence, these pulses can exhibit a chirp parameter larger than 12 $\frac{as}{eV}$ and thus  may fulfill condition  \textbf{(II)(b)}. For $\Delta E \sim$ 1 a.u., the effective time interval between core-excitation and dump transitions corresponding to chirp parameters $\beta$ between $12$ $\frac{as}{eV}$ and $41 \frac{as}{eV}$ ranges between $150$ as to $\sim600$ as so that these pulses created by the HHG process might be used to implement the approach presented. In particular, this might be useful to evidence and investigate charge migration in molecules upon core excitation which occurs on sub-fs timescales \cite{doi:10.1063/1.3506617}. \changeSecondSubmission{It should be noted, however, that the pulses created by the high-harmonic generation process exhibit chirp parameter with both positive as well as negative sign. In order to implement the approach presented, one therefore has to ensure that the atoms only interact with the part of the pulse that exhibits a positive chirp parameter $\beta$. This could be achieved by appropriatly selecting the long trajectories \cite{PhysRevLett.107.153902} or by phase-matching only the long trajectories.}

Also FEL pulses in the x-ray regime, which typically have a spectral bandwidth on the order of 1 $\%$ to several percent of the central photon energy \cite{RevModPhys.88.015007,ZAGORODNOV201669,1306.4830}, can cover the energy gap between ground state and valence-excited states (condition \textbf{(II)(a)}). Moreover, they can be chirped \cite{1367-2630-20-1-013010,Gauthier2016}. For instance at FLASH in Hamburg \cite{Ackermann2007}, the pulses could potentially be chirped in such a way that the arrival times of frequencies in the range up to $\sim 4\,\% $  of the photon energy are  stretched over a time interval on the order of tens of femtoseconds \cite{pcSMY}. This could allow one to trace dynamics on few 10-fs timescales which could be used to trace nuclear dynamics upon core-excitation.

\end{itemize}

\section{Application}

\begin{center}
\begin{table}[t]
\begin{tabular}{|c|c|c|}
  \hline
   \quad $|I\rangle$ \quad  & \quad  $d_{0,I} $(a.u.) \quad &  \quad $E_I$ (eV)\\\hline\multicolumn{3}{|c|}{$3s^23p^6\rightarrow2p^54s^1$}
  \\
\hline
   \quad $[2p^3_{3/2} 4s^1_{1/2}] P_1$ \quad  & \quad 0.0229 \quad &  \quad 244.0\\    \hline
   \quad $[2p^1_{1/2} 4s^1_{1/2}] P_1$ \quad  & \quad -0.0163 \quad &  \quad 246.1\\    \hline\hline
  \multicolumn{3}{|c|}{$3s^23p^6\rightarrow2p^53d^1$}
  \\
\hline
      \quad $[2p^3_{3/2} 3d^1_{5/2}] P_1$ \quad  & \quad -0.0239 \quad &  \quad 246.7\\    \hline
         \quad $[2p^1_{1/2} 3d^1_{3/2}] P_1$ \quad  & \quad 0.0169 \quad &  \quad 248.9\\    \hline
\end{tabular}
\caption{\label{tab:transition_dipoles_transition_energies_GS_CES}The dipole matrix elements $d_{0,I}$  as well as transition energies $E_I$  of the most relevant transitions between the ground state and the core-excited states $|I\rangle$ included in the model.\newline\newline }

\begin{tabular}{|c|c|c|c|}
  \hline
   \quad $|I\rangle$ \quad & \quad $|F\rangle$ \quad & \quad $d_{I,F}$ (a.u.)  \quad &  \quad $E_I$ - $E_F$ (eV) \quad\\
   \hline
  \hline
  \multicolumn{4}{|c|}{$2p^54s^1\rightarrow3s^14s^1$}
  \\
   \hline
   $ [2p^3_{3/2} 4s^1_{1/2}] \, P_1 $&\multirow{ 2}{*}{$  [3s^1_{1/2} 4s^1_{1/2}] \, S_1$} & -0.0618 &  214.7 \\
   $ [2p^1_{1/2} 4s^1_{1/2}] \, P_1  $ &\quad& -0.0831  & 216.8 \\
   \hline
 
   $ [2p^3_{3/2} 4s^1_{1/2}] \, P_1 $&\multirow{ 2}{*}{$  [3s^1_{1/2} 4s^1_{1/2}] \, S_0$} & -0.0852  & 214.3 \\
   $ [2p^1_{1/2} 4s^1_{1/2}] \, P_1 $ &\quad& 0.0603 & 216.4   \\
   \hline 
   \hline
  \multicolumn{4}{|c|}{$2p^53d^1\rightarrow3s^13d^1$}
  \\
   \hline

   $ [2p^3_{3/2} 3d^1_{5/2}] \, P_1 $&\multirow{ 2}{*}{$  [3s^1_{1/2} 4d^1_{3/2}] \, S_2$} & 0.06 & 214.8 \\
   $ [2p^1_{1/2} 3d^1_{3/2}] \, P_1  $ &\quad& 0.0674  & 217.0 \\
   \hline 
 
   $ [2p^3_{3/2} 3d^1_{5/2}] \, P_1 $&\multirow{ 2}{*}{$  [3s^1_{1/2} 4d^1_{5/2}] \, S_2$} & -0.0809  & 214.8 \\
   $ [2p^1_{1/2} 3d^1_{3/2}] \, P_1 $ &\quad& 0.0639 & 216.9    \\
   \hline 
\end{tabular}

\caption{\label{tab:transition_dipoles_transition_energies_CES_VES}The non-vanishing dipole matrix elements $d_{I,F}$ in atomic units as well as transition energies ($E_I$ - $E_F$) in eV of the most relevant transitions between the core-excited states $|I\rangle$ and valence excited states $|F\rangle$ included in the model.}
\end{table}
\end{center}

%

\begin{figure*}
\centering
\subfloat[][approximate theory (Eq. \ref{eq:difference_spectrum}). Propagation effects are not taken into account.]{\includegraphics[width=0.40\textwidth]{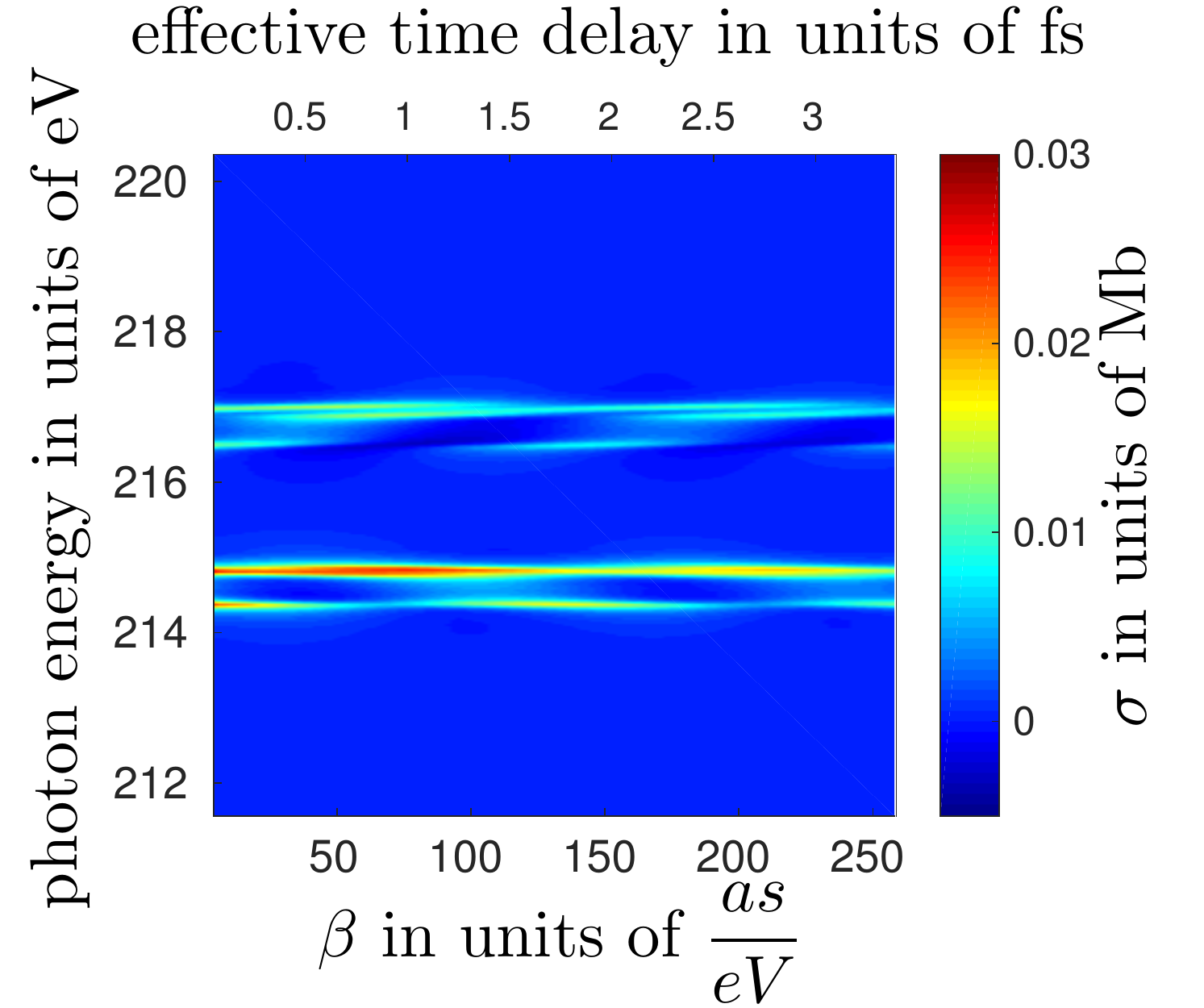}}
\subfloat[][numerical simulation. $\rho L: 0.01 \times 10^{18}\,$cm$^{-2}$]{\includegraphics[width=0.40\textwidth]{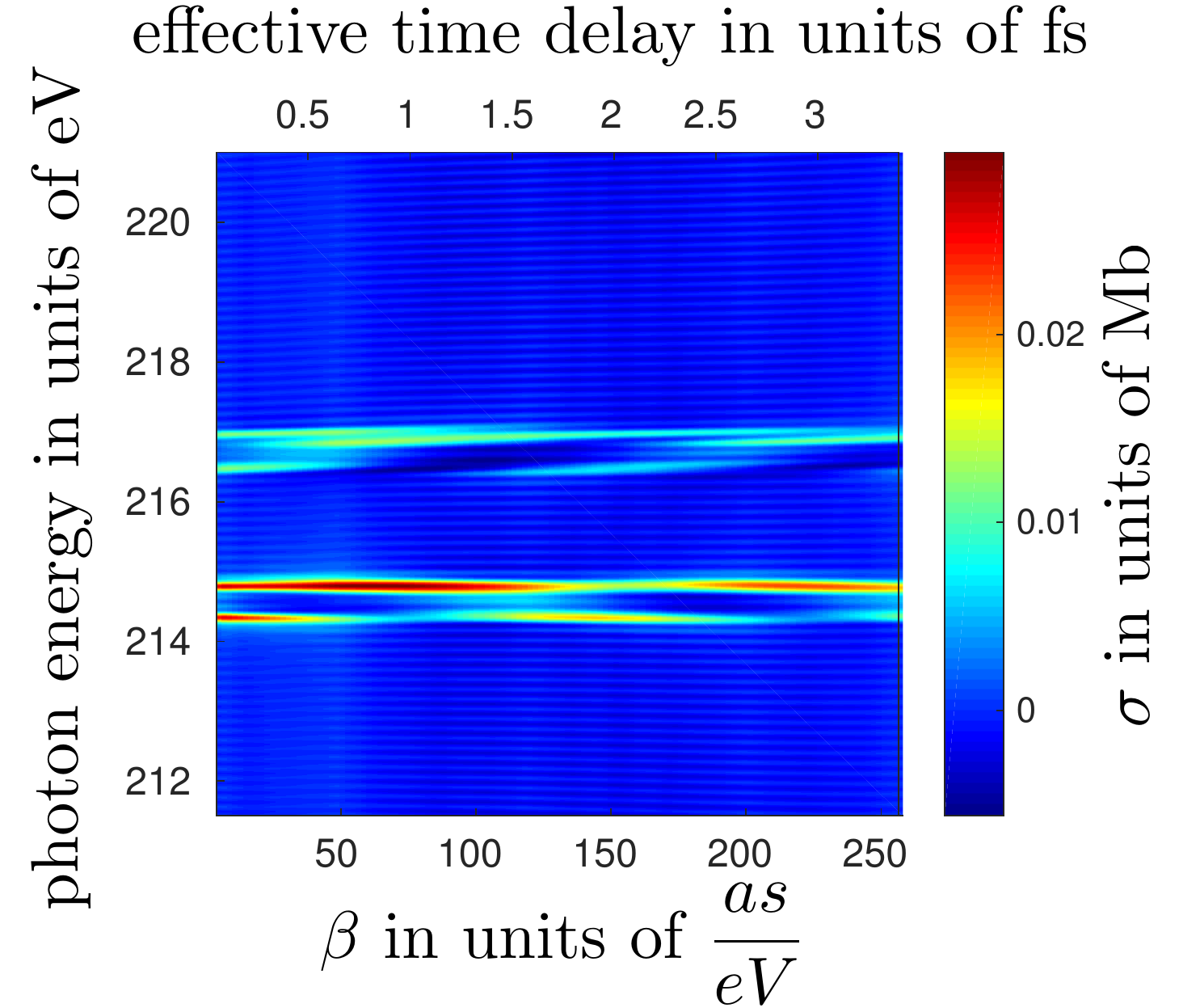}}
\\
\subfloat[][numerical simulation. $\rho L: 0.5 \times 10^{18}\,$cm$^{-2}$]{\includegraphics[width=0.40\textwidth]{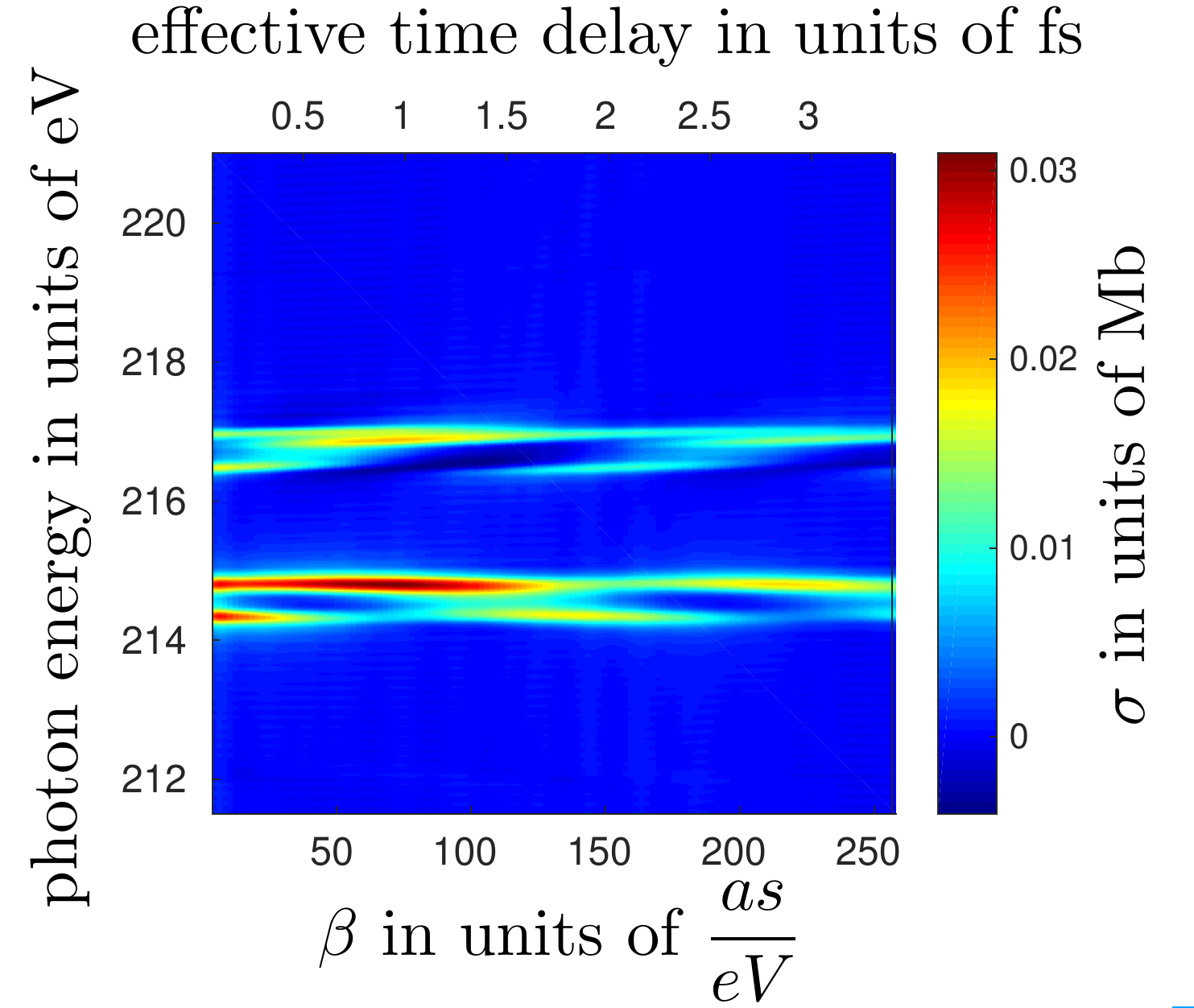}}
\subfloat[][numerical simulation. $\rho L: 1 \times 10^{18}\,$cm$^{-2}$]{\includegraphics[width=0.40\textwidth]{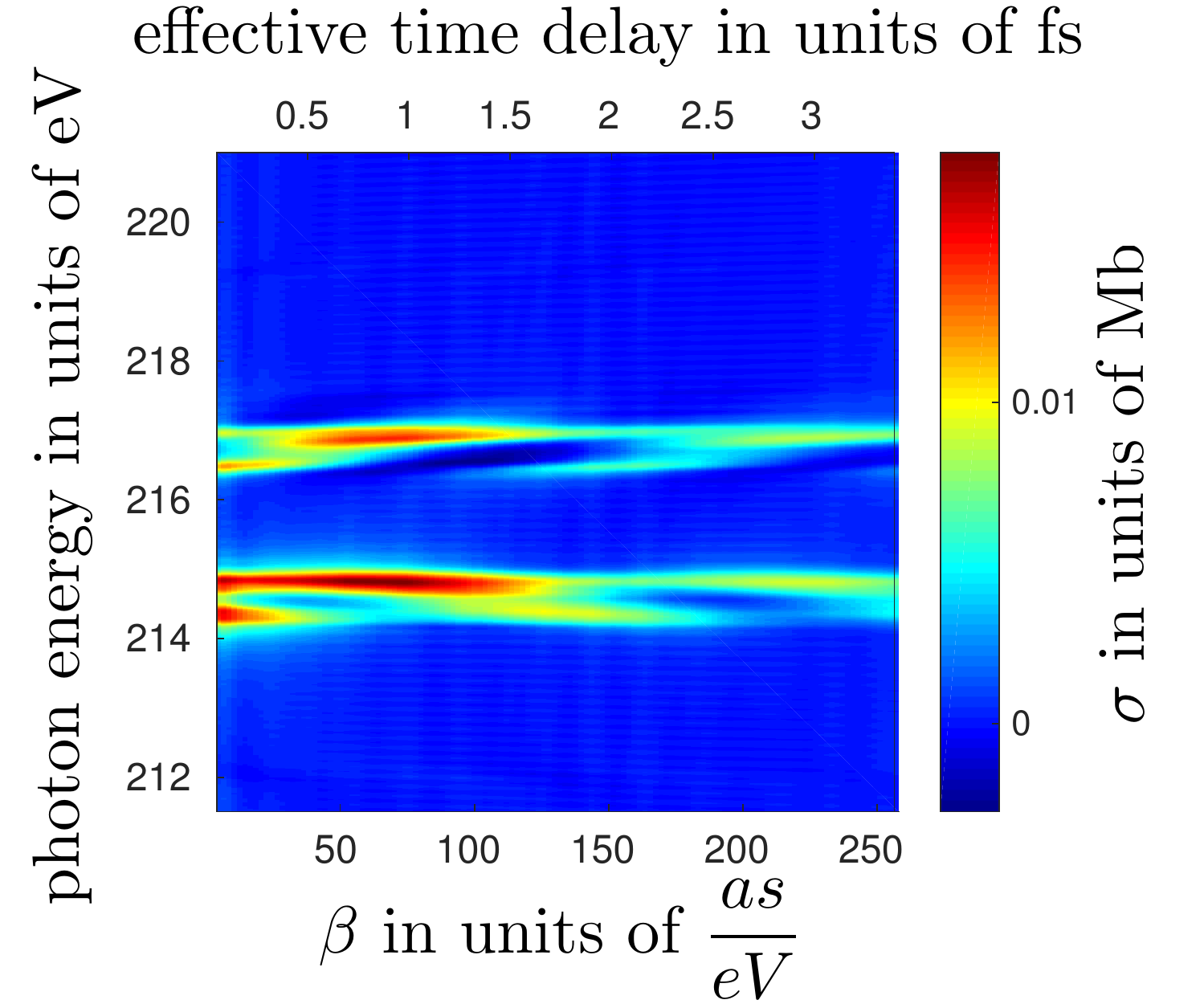}}
  \caption{\label{fig:chirpdependent_spectrum_tdse_and_approximation}The emission cross section (a) $\sigma$ (see Eq. \ref{eq:difference_spectrum}) according to the approximate theory (Eq. \ref{eq:difference_spectrum}).  (b-d) $\sigma_{app}$ (see Eq. \ref{eq:sigmaapp}) obtained from  numerical simulations using pulses parametrized as described in the text with a carrier frequency of 230 eV, 10 nJ pulse energy and 500 nm focal radius. In the numerical simulations, the field was propagated for various distances $L$ so that density-length products $\rho L =$ 0.01,0.5 and 1 $\times$ 10$^{18}$cm$^{-2}$ were reached.}

\end{figure*}

In the following, we exemplify the approach by means of SRIXS at the argon L-edge. For the simulation of SRIXS, we solved the coupled Schr\"odinger equation and Maxwell's equations numerically in the rotating wave approximation and the slowly varying envelope approximation as described in \cite{PhysRevA.90.063828}.  The atomic transition dipole matrix elements and the electronic energies were obtained from the relativistic configuration interaction code FAC \cite{FACcode} whereby the calculations were restricted to dipole-allowed transitions involving states in which electrons populate the shell with main quantum numbers from 1 up to 4. Valence ionization as well as the Auger decay of core holes is phenomenologically included as described in Ref. \cite{PhysRevA.90.063828}. \changeSecondSubmission{ The decay rates of the core-excited states considered, which were needed for the numerical simulations as well as for the analytical emission cross section, were determined by the flexible atomic code (FAC) \cite{FACcode} to be approximately 100 meV. The valence-ionization cross section was set, according to \cite{YEH19851}, to $\sim 0.4$ Mb.} The envelope of the fields considered were chosen to be of Gaussian form (i.e., $\mathcal{E}(t)= \mathcal{E}_{max}e^{-\frac{t^2}{2\tau^2}}$ ); the amplitude $\mathcal{E}_{max} = \frac{E_0}{(1+n^2)^{\frac{1}{4}}}$, the pulse duration $\tau = \sqrt{1+n^2}$ and the chirp parameter $\beta = \frac{2(1+n^2)}{n}$ were parametrized such that, as in Refs. \cite{PhysRevA.84.013417,PhysRevA.95.043424}, the power spectra of the pulses are chirp parameter independent. In the calculations, the parameter $n$ was varied between 1 and 128.

We chose the central carrier frequency $\omega_0$ 
to 230 eV and the spectral bandwidth was set to be $\sim 45$ eV  so that the pulses employed are able to induce transitions from both the ground state to the core excited states that are associated with the promotion of a 2p electron to the $n$d ($n\geq3$) or $n$s ($n\geq4$) orbitals as well as from these respective core excited states to the valence excited states that are reached by the refilling of the core hole by a 3s electron.

\begin{figure}
\includegraphics[width=1\linewidth]{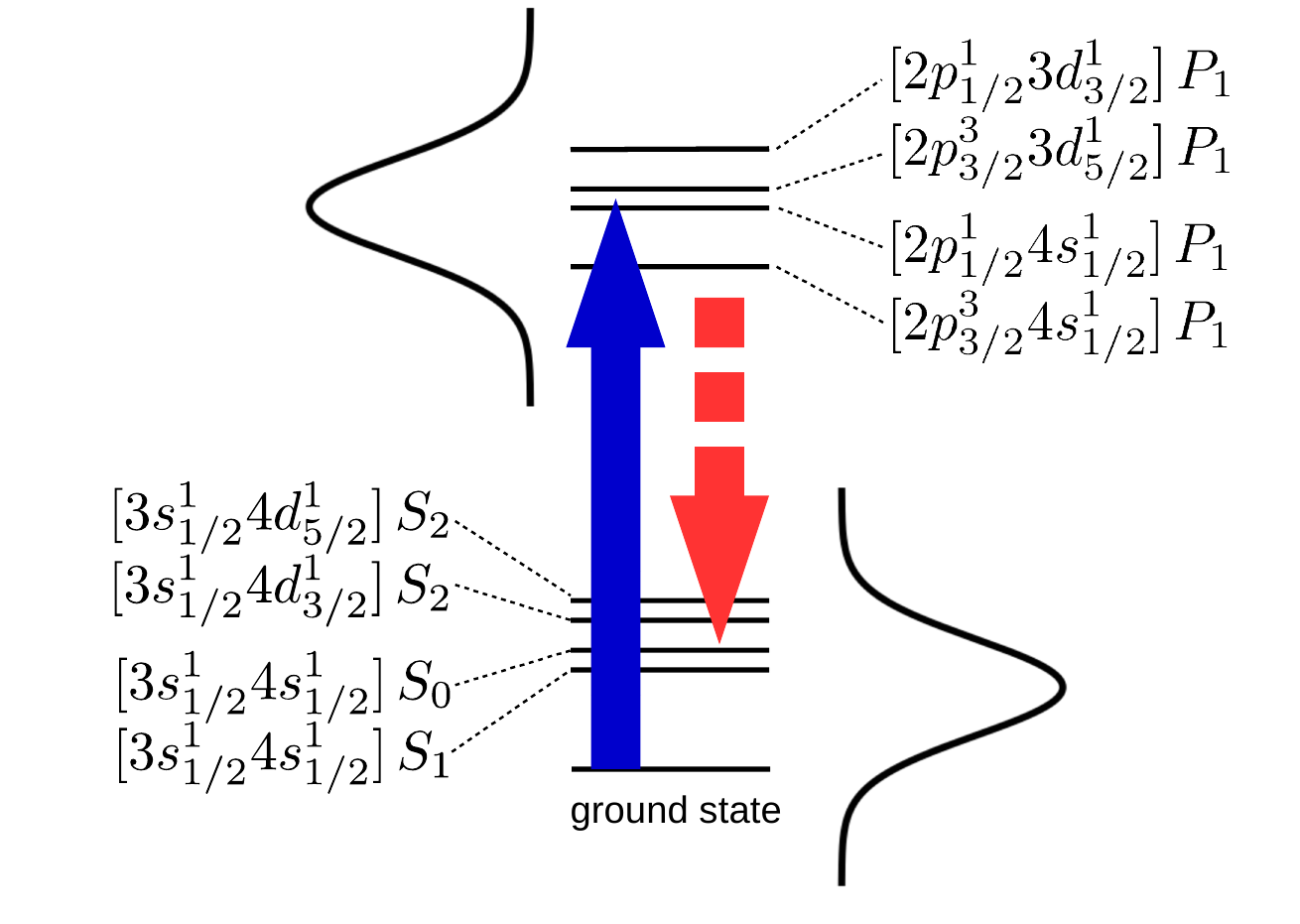}

\caption{\label{fig:level_scheme}The most relevant level involved in the dynamics induced by the chirped broadband pulses considered in the numerical simulations.
}
\end{figure}

\begin{figure}
\includegraphics[width=1\linewidth]{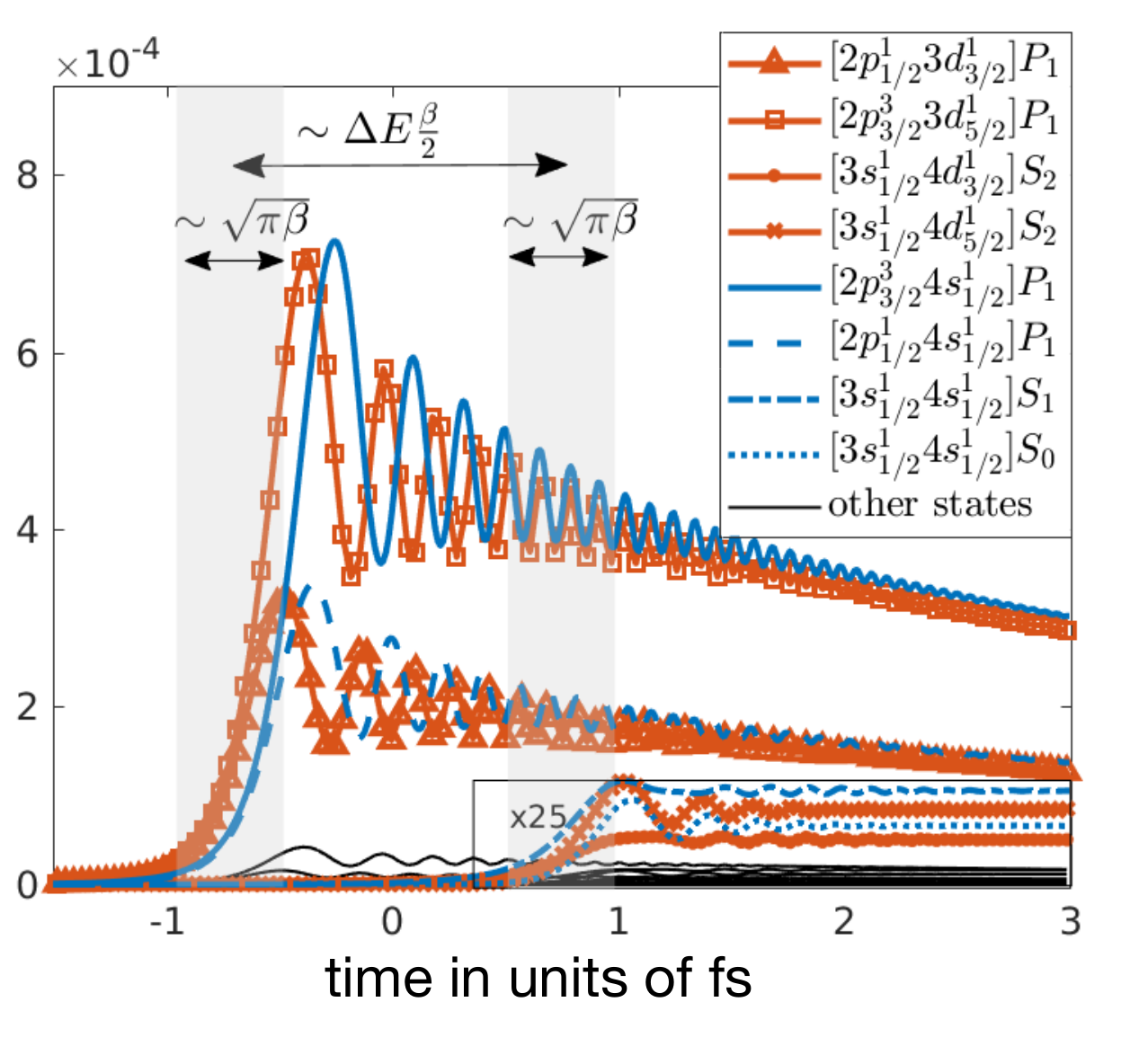}

\caption{\label{coefficients_n_80}
The time-dependent populations of core-excited states and valence excited states caused by the interaction with a chirped pulse with electric field parametrized as described in the text with a carrier frequency of 230 eV, 10 nJ pulse energy, 500 nm focal radius and chirp parameter $\beta = 90\frac{as}{eV}$. }
\end{figure}

For the fields considered, one finds  that the pulses  at first primarily create a coherent superposition of the core-excited states that are associated with the the promotion of a 2p electron to a 4s orbital (i.e., the spin-orbit-split states :$[2p^3_{3/2} 4s^1_{1/2}] \, P_1$ and $ [2p^1_{1/2} 4s^1_{1/2}] \, P_1$) and to a 3d orbital (i.e.,  $[2p^3_{3/2} 3d^1_{5/2}] \, P_1$ and $[2p^1_{1/2} 3d^1_{3/2}] \, P_1$).        This is shown in Fig. \ref{coefficients_n_80} for a pulse with chirp parameter $\beta = 90\frac{as}{eV}$. The excitation process is essentially restricted to a time interval of length $\sqrt{\pi\beta}$ which is approximately 400 as for $\beta = 90\frac{as}{eV}$. 
 Subsequently, the populations of the core-excited states exhibit oscillations with decreasing amplitude around a decreasing mean value whereby the  decreasing mean-value reflects the Auger decay of the 2p core holes with a life time of $\sim 5.7$ fs \cite{CARROLL200167}.
 The second interaction of the pulses induces then transitions that are associated with the refilling of the 2p hole by a 3s electron.
In Fig. \ref{coefficients_n_80}, this can be seen by the increasing populations of valence-excited states $\sim$ 2 fs after the core-excitations. Also these dump transitions occur within a time interval with a length on the order of a few hundred attoseconds. 

The light-matter interaction associated with the dump-transitions results in chirp-dependent (i.e. effective-time-delay-dependent) features in the spectrum of the transmitted light at 214 - 216 eV photon energy. This is shown in Fig. \ref{fig:chirpdependent_spectrum_tdse_and_approximation} b)-d) where the chirp-dependent normalized apparent emission cross section $\sigma_{app}$ (Eq. \ref{eq:sigmaapp}), a quantity obtainable from the numerical simulations that can be directly compared to the cross section $\sigma$ obtained from the approximate theory (Eq. \ref{eq:difference_spectrum}), is depicted for various propagation distances at photon energies in the range of the dump-transition energies. 

\begin{equation}\label{eq:sigmaapp}\sigma_{app} = \frac{1}{\rho L}\ln\left(\Big|\frac{\tilde{\mathcal{E}}(0,\omega_0)}{\tilde{\mathcal{E}}(L,\omega_0)}\Big|^{2}\right)\ln\left(\Big|\frac{\tilde{\mathcal{E}}(L,\omega)}{\tilde{\mathcal{E}}(0,\omega)}\Big|^{2}\right)\end{equation} 

The features in the cross section can be related to the transitions given in Tab. \ref{tab:transition_dipoles_transition_energies_CES_VES}. They split in two groups with transition energies 214-215 eV and 216-217 eV, respectively, which are reflected in two groups of lines in the  emission cross section shown in Fig. \ref{fig:chirpdependent_spectrum_tdse_and_approximation}. Unaffected by the propagation distance, the lines exhibit a characteristic beating for varying chirp parameter $\beta$ with a period of $\sim 130\frac{as}{eV}$ corresponding to an effective time-delay difference of $\sim 2$fs. The increase of the propagation distance and the associated increasing impact of propagation effects, predominantly cause a broadening of the lines and affect the amplitudes of the features (see Fig. \ref{fig:chirpdependent_spectrum_tdse_and_approximation}b - d) but they do not affect the beating period. In particular for short propagation distances (e.g. for a density length product of $\rho L = 0.01 \times 10^{18}$cm$^{-2}$) where the impact of propagation effects is reduced with respect to large propagation distances, the features and in particular the characteristic beating are accurately reproduced by the approximate theory developed (see Fig. \ref{fig:chirpdependent_spectrum_tdse_and_approximation}a and b). Therefore, the oscillatory behavior of the features can be related on the basis of Eq. \ref{eq:difference_spectrum} to the existence of coherences of the time-evolved core-excited state launched by core excitation. 
Considering that, according to Eq. \ref{eq:difference_spectrum}, only coherences between core excited states that are coupled to the same valence excited state can be reflected in the spectra, the beating can be related to the coherences between $2p^54s^1$-spin-orbit-split states as well as between $2p^53d^1$-spin-orbit-split states (see Tab. \ref{tab:transition_dipoles_transition_energies_CES_VES} ). These levels are split by $2.1 $eV and $2.2 $eV, respectively, corresponding to dynamical timescales on the order of 2 fs which is consistent with the period of the beating in the chirp-dependent emission cross section. 

\begin{figure}
\centering
\subfloat[][pulse energy: 10 nJ]{\includegraphics[width=0.24\textwidth]{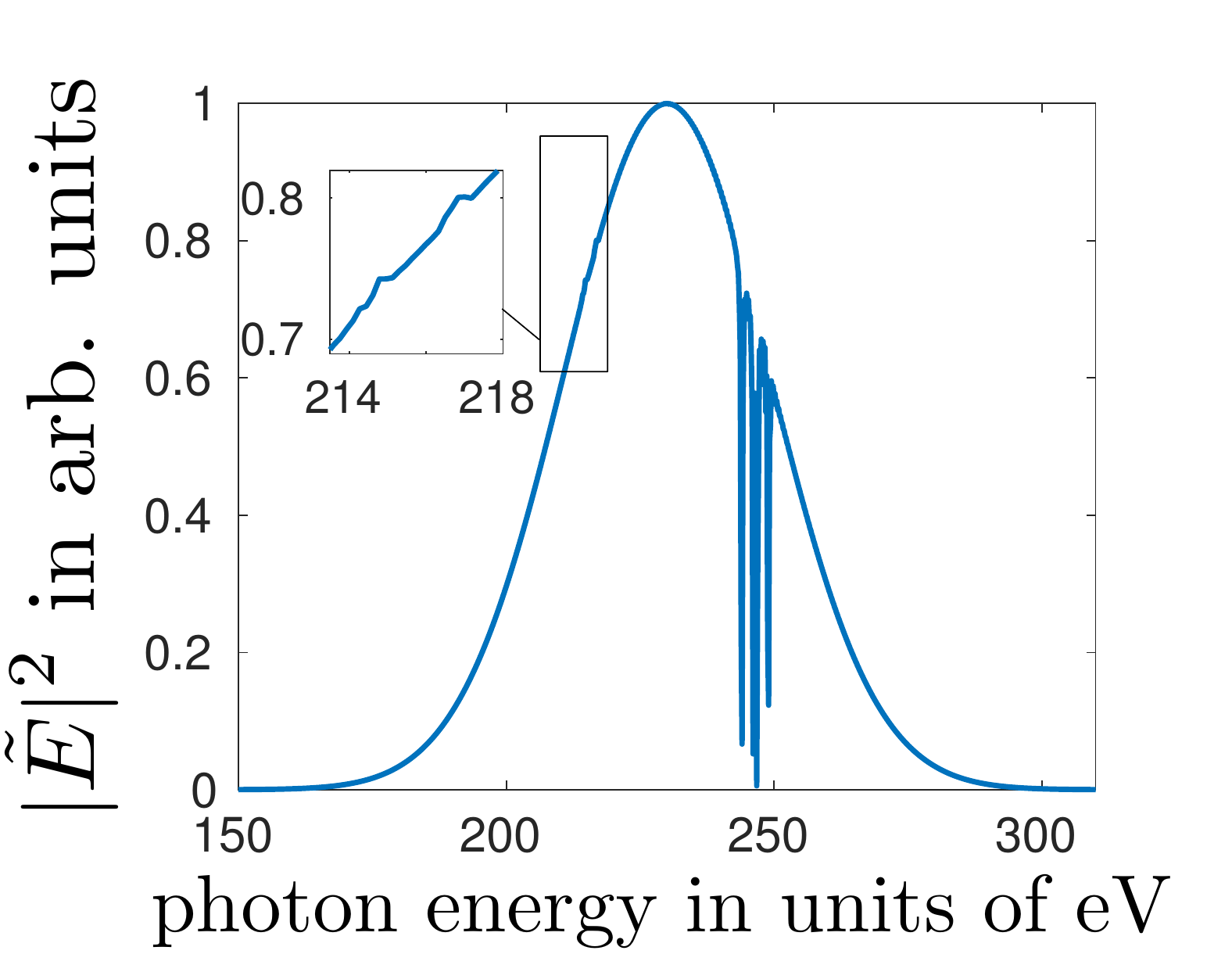}}
\subfloat[][pulse energy: 100 nJ]{\includegraphics[width=0.24\textwidth]{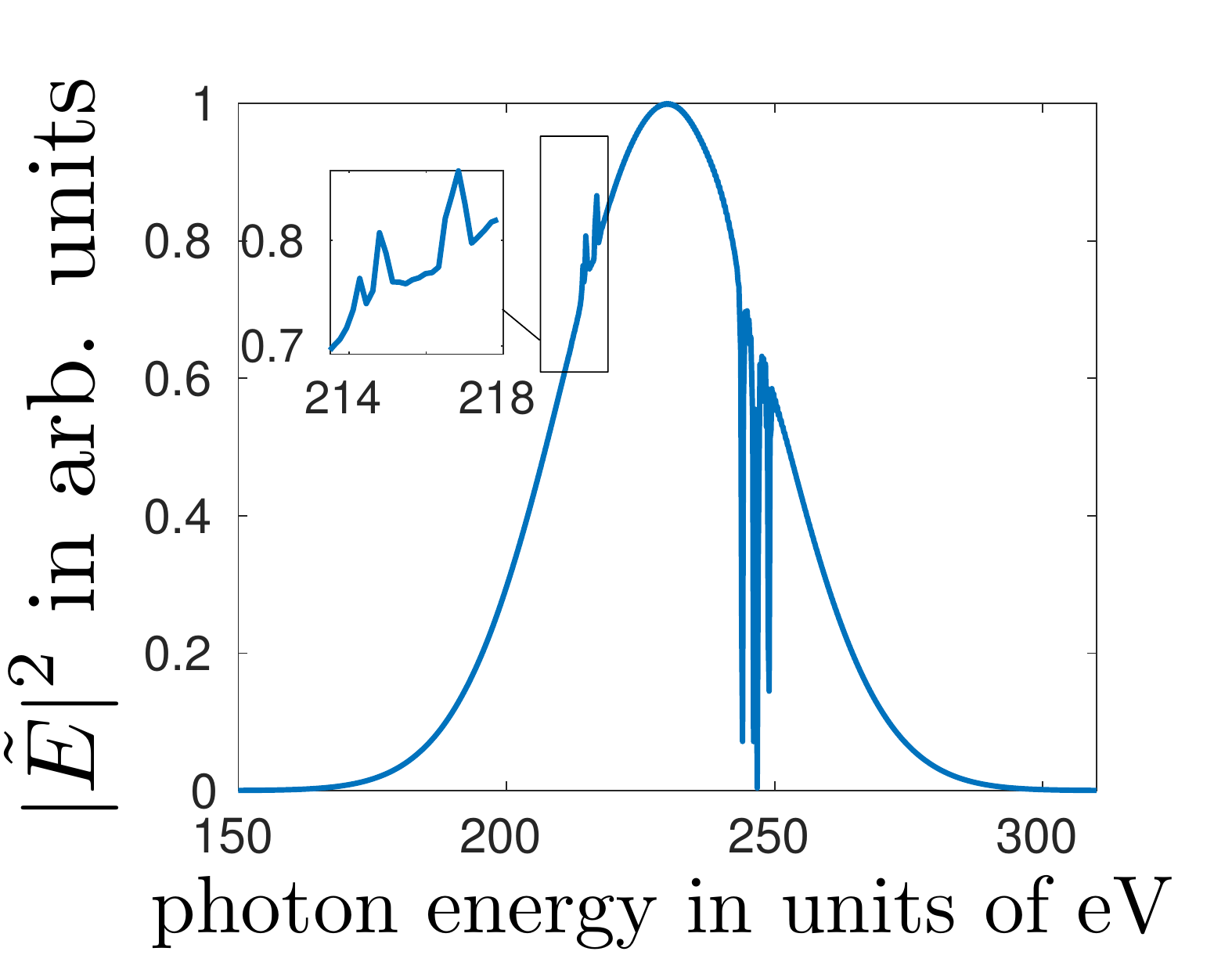}}
 \qquad
\subfloat[][pulse energy: 1 $\mu$J]{\includegraphics[width=0.24\textwidth]{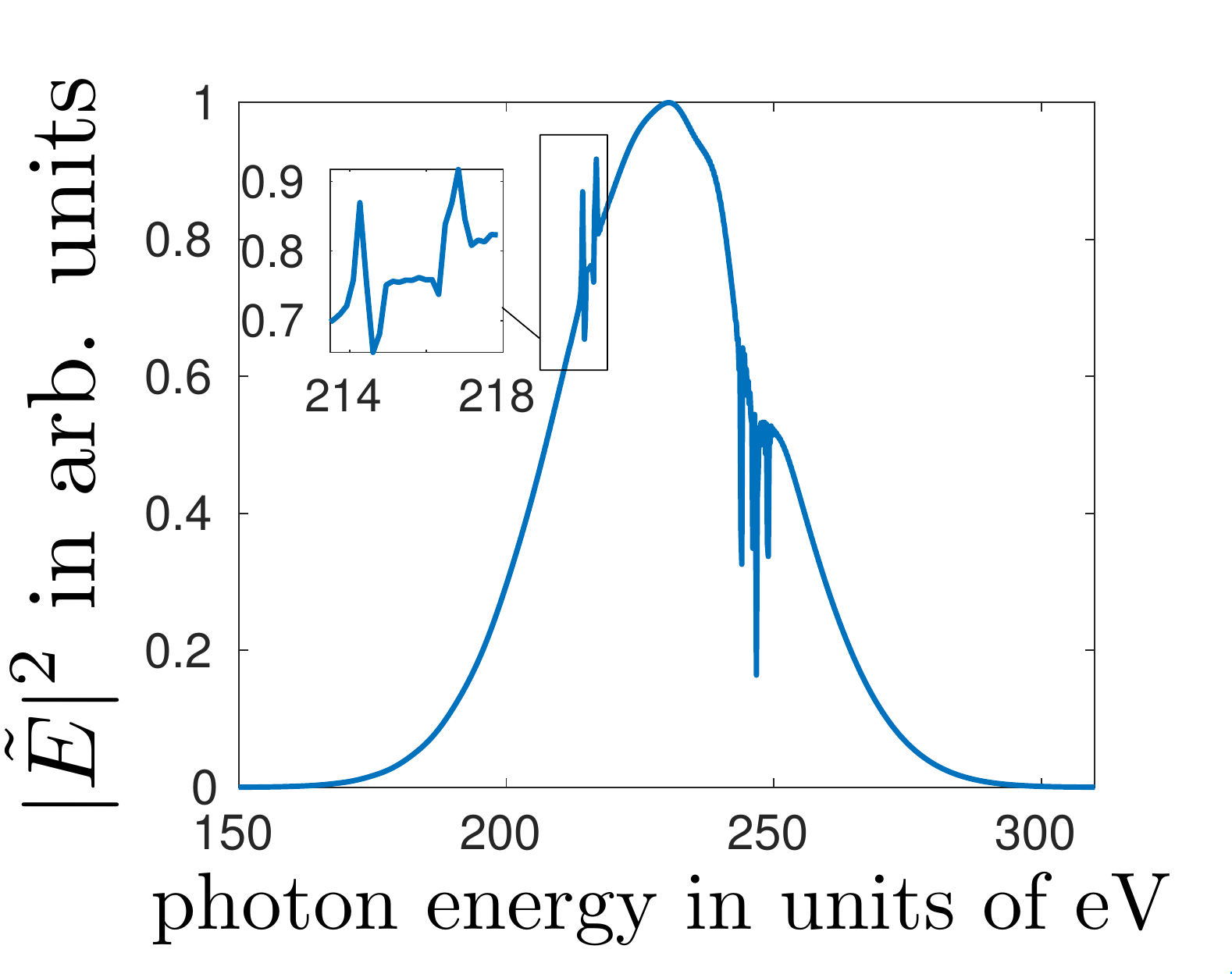}}
  \caption{\label{fig:spectrum_line_plot}The spectrum of the transmitted field of a chirped pulse with chirp parameter $\beta=35\frac{as}{eV}$ parametrized as described in the text with a carrier frequency of 230 eV, 10 nJ pulse energy and 500 nm focal radius. Significant features  associated with the SRIXS process (see inset of the plots) can be observed for pulses with pulse energies in the nJ regime. For 10 nJ, which correspond to $10^7$ photons/eV at 215 eV, a modulation of the spectrum on the order of $1\%$ can be observed. For 1 $\mu$J, the modulation of the spectrum is on the order of 30 $\%$. The  fields were propagated in space until a density-length product of $1\times10^{-18}$cm$^{-2}$ was reached.}
\label{fig:cont}

\end{figure}

Finally, we point out that with the attosecond pulses available \cite{Ferrari2010,Takahashi2013,Sansone2011,PhysRevLett.120.093002}, when focussed to a focal diameter on the order of 1 $\mu m$, a significant SRIXS signal can indeed be induced causing a modulation of the spectra of the transmitted light on the order of 1 $\%$ to several tens of percent at 10 nJ and 1$\mu$J pulse energies respectively (see Fig. \ref{fig:spectrum_line_plot}). Hence, an implementation of the technique presented with the ultra-short pulses available appears to be feasible. \newline

%

\section{Conclusions}
To conclude, we presented an approach for initiating and tracing ultra-fast electron dynamics in core-excited atoms, molecules or solids which is based on stimulated resonant inelastic x-ray scattering induced by single, chirped, broadband XUV/x-ray pulses. The approach can be implemented with available ultra-short pulses providing insight into core-excited state dynamics. With respect to conventional pump-probe experiments that use two pulses, the approach presented offers the advantages that it allows one to implement a pump-probe experiment with a single pulse which does not need the synchronization of two pulses. Moreover, the time-resolution achievable is not limited by the pulse duration---it is only limited by the chirp. It thus provides sub-pulse-duration time resolution and might therefore be used to push the achievable time resolution at table-top sources as well as FELs towards shorter timescales.
\newline
\begin{acknowledgments}\change{
The authors are grateful to  Evgeny Schneidmiller, Mikhail Yurkov, Svitozar Serkez and Gianluca Geloni for clarifying discussions concerning the chirp of the pulses delivered by FLASH and the European XFEL in Hamburg. }
\end{acknowledgments}
\appendix* \section{}In the following, we derive the approximate expressions for the transition amplitudes $\alpha_I(\tau)$ and $\alpha_F(\tau)$ given in the main text. Concretely, we approximate them by the respective first non-vanishing term of the perturbation series.
\subsection{Transition amplitudes of core-excited states  $\alpha_I$}
 Assuming that prior to the exposure to the field, the atoms are in their ground state $|0\rangle$, we approximate the coefficients $\alpha_I(\tau)$ by the first order term:

\begin{equation}
\alpha_I(\tau)\sim
-\frac{1}{2i}\int_{-\infty}^{\tau}dte^{i(E_I-E_0)t}d_{I,0}\big[\mathcal{\overline{E}}(t)e^{-i(\omega_{0}t-\frac{t^{2}}{\beta})}+c.c.\big]
\end{equation}
where $d_{I,0}$ denote the dipole transition matrix elements between ground state and core-excited states. In the rotating wave approximation, this simplifies to:
\begin{equation}\sim-\frac{1}{2i}\int_{-\infty}^{\tau}dte^{i(E_I-E_0)t}d_{I,0}\mathcal{\overline{E}}(t)e^{-i(\omega_{0}t-\frac{t^{2}}{\beta})}\end{equation}
Noting that the integrand oscillates quickly for $|t - t_I| \gtrsim \sqrt{\beta} $ with  $t_I = (\omega_0 - E_I + E_0) \frac{\beta}{2} $, predominantly values in the vicinity of $t_I$ contribute to the integral so that, given that the amplitude $\mathcal{\overline{E}}(t)$ varyies neglegibly within the time interval $[t_I - \sqrt{\beta},t_I + \sqrt{\beta}]$, one may approximate the above expression by:

 \begin{equation}\alpha_I(\tau)\sim-\frac{1}{2i}d_{I,0}\mathcal{\overline{E}}(t_I)e^{ - i\frac{t^2_I}{\beta}}\int_{-\infty}^{\tau}dte^{i\frac{(t - t_I)^2}{\beta}}.\end{equation}

 For $\tau \gg t_I +\sqrt{\beta}$, the integral becomes mostly $\tau$-independent so that for this limiting case, the upper border of the integral might be set to $+\infty$ which allows one to evalueate the integral analytically. With this, one finds for the transition ampitude $\alpha_I(\tau \gg t_I +\sqrt{\beta})$: 
 \begin{equation}  \label{eq:alphaI}\alpha_I(\tau \gg t_I +\sqrt{\beta}) \sim-\frac{1}{2i}d_{I,0}\mathcal{\overline{E}}(t_I)e^{ - i\frac{t^2_I}{\beta}}\sqrt{\pi\beta}e^{i\frac{\pi}{4}}.\end{equation}

\subsection{Transition amplitudes to valence-excited states  $\alpha_F$}
Turning to the transition amplitude of the valence-excited states $|F\rangle$, the first non-vanishing term in the perturbation series is the second order one: \begin{equation}
\alpha_F(\tau)\sim \frac{1}{i}\sum_{I}\int_{-\infty}^{\tau}dt\alpha_{I}(t)\langle F|\hat{V}(t)|I\rangle
 \end{equation}
\begin{equation}
=-\frac{1}{i}\sum_{I}\int_{-\infty}^{\tau}dt\alpha_{I}(t)d_{F,I}e^{i(E_F-E_I)t}\frac{1}{2}\big[\mathcal{\overline{E}}(t)e^{-i(\omega_{0}t-\frac{t^{2}}{\beta})}+c.c.\big]
\end{equation}
In the rotating wave approximation, this simplifies to:
\begin{equation}
\sim-\frac{1}{i}\sum_{I}\int_{-\infty}^{\tau}dt\alpha_{I}(t)d_{F,I}e^{i(E_F-E_I)t}\frac{1}{2}\mathcal{E}(t)e^{i(\omega_{0}t-\frac{t^{2}}{\beta})}
\end{equation}
Noting that the integrand oscillates quickly for $|t - t_{F,I}| \gtrsim \sqrt{\beta} $ with $t \sim t_{F,I} = (E_F - E_I + \omega_0)\frac{\beta}{2} $, predominantly values in the vicinity of $t \sim t_{F,I}$ contribute to the integral. If $|t_{I} - t_{F,I}| \gg \sqrt{\pi \beta} $, the coefficients $\alpha_{I}(t)$ can be assumed to be time independent. With this, one can find an approximation for the second-order coefficients for this particular limiting case using analogous steps as described above for the transition amplitudes $\alpha_I(\tau \gg t_I +\sqrt{\beta})$: 
\begin{multline}
\alpha_F(\tau \gg t_{F,I}+\sqrt{\beta}) \sim -\frac{1}{2i}\sum_{I}
\alpha_I(t_{F,I})\\
\mathcal{E}(t_{F,I})
d_{F,I} e^{i\frac{t_{F,I}^{2}}{\beta}}\sqrt{\pi\beta}e^{-i\frac{\pi}{4}}
\end{multline}

\subsection{Spectrum of the transmitted light}
In the following, we will derive an approximate expression for the spectrum of an x-ray pulse propagating through a sample of atoms at photon energies in the vicinity of the dump-transition energies. Eventuelly, this will allow us to determine an approximate expression for the emission cross section in this spectral region. Here, we will consider the impulsive limit, i.e. the situation where the temporal range spanned by the times $t_I$ and the times $t_{F,I}$, respectively, is small in comparison to the dynamical timescales being probed, small in comparison to $\sqrt{\beta}$ as well as small in comparison to the characteristic timescale on which the slowly varying envelope $\mathcal{E}$ changes. Moreover, propagation effects will not be taken into account, i.e., we assume that the field-induced transition amplitudes remain independent of the propagation distance.

Our procedure will be devided in two steps: First, we will determine the Fourier transform of the polarization that is associated with the field-induced dump-transitions. 

In a second step, we will use this expression, which enters the right side of Eq. \ref{eq:mwe_SVEA2}, to obtain the spatial evolution of the spectral amplitude of the electric field in the spectral region associated with the dump-transition energies.
 \begin{equation}\label{eq:mwe_SVEA2}
\frac{\partial\tilde{\mathcal{E}}(z,\omega)}{\partial z} = 2\pi i\frac{\omega}{c}\tilde{\mathcal{P}}(z,\omega).
 \end{equation} 
   
\subsubsection{Field-induced polarization}
The polarization associated with the dump transitions $\mathcal{P}_{Dump}(z,t)$ is given by:
\begin{equation}\mathcal{P}_{Dump}(z,t)=\rho\sum_{I,F} \overline{\alpha_I}(t)\alpha_F(t)\langle I|D|F\rangle + c.c.\end{equation}
Here, $\rho$ denotes the atomic number density and $D$ is the dipole operator. In the impulsive limit, the coefficients may be approximated by:
\begin{equation}
\alpha_{I}(t)\sim\Theta(t-t_{I})\alpha_{I}(t\gg t_{I}+\sqrt{\beta})e^{-\frac{\Gamma_I (t-t_I)}{2}}
\end{equation}
\begin{multline}
\alpha_{F}(t)\sim-\frac{1}{2i}\sum_{I}\Theta(t-t_{F,I})
\alpha_I(t\gg t_{I}+\sqrt{\beta})e^{-\frac{\Gamma_I (t_{F,I}-t_I)}{2}} \\
\mathcal{E}(t_{F,I})
d_{F,I} e^{i\frac{t_{F,I}^{2}}{\beta}}\sqrt{\pi\beta}e^{-i\frac{\pi}{4}}
\end{multline} where phenomenologically, exponential factors $e^{-\frac{\Gamma_I (t-t_I)}{2}}$ and $e^{-\frac{\Gamma_I (t_{F,I}-t_I)}{2}}$, respectively,  are  introduced to take into account the finite life times $\frac{1}{\Gamma_I}$ of the core-excited states $|I\rangle$. With this, the polarization $\mathcal{P}_{Dump}(z,t)$ may be approximated by:
 
\begin{multline}\label{eq:pdump_approx}
\mathcal{P}_{Dump}(z,t) =-\frac{\rho}{2i}  \sum_{I,I',F} \Theta(t-t_{F,I'})\overline{\alpha_I}(t\gg t_{I}+\sqrt{\beta})\\e^{-\frac{\Gamma_I (t-t_I)}{2}}
\alpha_{I'}(t\gg t_{I'}+\sqrt{\beta}) e^{-\frac{\Gamma_{I'} (t_{F,I'}-t_{I'})}{2}} \mathcal{E}(t_{F,I'}) \\d_{F,I'}e^{i\frac{t_{F,I'}^{2}}{\beta}}\sqrt{\pi\beta}e^{-i\frac{\pi}{4}} d_{I,F}e^{i(E_I - E_F)t} + c.c.
\end{multline}. 

In the following, we will derive an approximate expression of the Fourier transform of $\mathcal{P}_{Dump}(z,t)$.

The Fourier transform of $\mathcal{P}_{Dump}(z,t)$ is given by:
\begin{equation}
\tilde{\mathcal{P}}_{Dump}(z,\omega)=\int dte^{i\omega t}\mathcal{P}(z,t)
 \end{equation}
 which is, using Eq. \ref{eq:pdump_approx}, approximately:
\begin{multline}
=-\frac{\rho}{2i}\sum_{I,I',F} \overline{\alpha_I}(t\gg t_{I}+\sqrt{\beta})\alpha_{I'}(t\gg t_{I'}+\sqrt{\beta})
\\\mathcal{E}(t_{F,I'}) e^{i\frac{t_{F,I'}^{2}}{\beta}}\sqrt{\pi\beta}e^{-i\frac{\pi}{4}}d_{F,I'}d_{I,F} \\ i\frac{e^{i(E_I-E_F + \omega)t_{F,I'}}}{E_I-E_F + \omega + i\frac{\Gamma_I}{2}}e^{-\frac{\Gamma_I}{2} (t_{F,I'} - t_{I})} e^{-\frac{\Gamma_{I'}}{2} (t_{F,I'} - t_{I'})} \\+\frac{\rho}{2i}\sum_{I,I',F}\overline{\alpha_{I'}}(t\gg t_{I'}+\sqrt{\beta})\alpha_{I}(t\gg t_{I}+\sqrt{\beta})\\\overline{\mathcal{E}}(t_{F,I'}) e^{-i\frac{t_{F,I'}^{2}}{\beta}}\sqrt{\pi\beta}e^{i\frac{\pi}{4}}d_{I',F}d_{F,I}\\i\frac{e^{i(E_F-E_I +\omega)t_{F,I'}}}{E_F-E_I + \omega +\frac{i\Gamma_I}{2}} e^{-\frac{\Gamma_I (t_{F,I'} - t_I)}{2}}e^{-\frac{\Gamma_{I'} (t_{F,I'} - t_{I'})}{2}}
\end{multline}
For $\omega > 0$ one may neglect terms that are proportional to:\begin{equation}
\frac{1}{E_I-E_F + \omega + i\frac{\Gamma_I}{2}}
\end{equation}
so that:
\begin{multline}
\tilde{\mathcal{P}}_{Dump}(z,\omega > 0)\sim
\\\frac{\rho}{2}\sum_{I,I',F}\overline{\alpha_{I'}}(t\gg t_{I'}+\sqrt{\beta})\alpha_{I}(t\gg t_{I}+\sqrt{\beta})\\\overline{\mathcal{E}}(t_{F,I'}) e^{-i\frac{t_{F,I'}^{2}}{\beta}}\sqrt{\pi\beta}e^{i\frac{\pi}{4}}d_{I',F}d_{F,I}\\\frac{e^{i(E_F-E_I +\omega)t_{F,I'}}}{E_F-E_I + \omega +\frac{i\Gamma_I}{2}} e^{-\frac{\Gamma_I (t_{F,I'} - t_I)}{2}}e^{-\frac{\Gamma_{I'} (t_{F,I'} - t_{I'})}{2}}.
\end{multline}
Using Eq. \ref{eq:alphaI}, one finds:
\begin{multline}\label{eq:pdump_a}
\tilde{\mathcal{P}}_{Dump}(z,\omega > 0)\sim
\\\frac{\rho\pi\beta}{8}\sum_{I,I',F}d_{0,I'}\mathcal{E}(t_{I'})e^{  i\frac{t^2_{I'}}{\beta}}
d_{I,0}\mathcal{\overline{E}}(t_I)e^{ - i\frac{t^2_I}{\beta}}\\\overline{\mathcal{E}}(t_{F,I'}) e^{-i\frac{t_{F,I'}^{2}}{\beta}}\sqrt{\pi\beta}e^{i\frac{\pi}{4}}d_{I',F}d_{F,I}\\\frac{e^{i(E_F-E_I +\omega)t_{F,I'}}}{E_F-E_I + \omega +\frac{i\Gamma_I}{2}} e^{-\frac{\Gamma_I (t_{F,I'} - t_I)}{2}}e^{-\frac{\Gamma_{I'} (t_{F,I'} - t_{I'})}{2}}
\end{multline}
Focussing on the prominent features which occur at frequencies \begin{equation}\omega \sim E_I - E_F\end{equation}
one may set
\begin{equation}e^{i(E_F-E_I +\omega)t_{F,I'}} \sim 1,
.\end{equation} In the impulsive limit, it holds true that
\begin{equation}\mathcal{\overline{E}}(t_I)\sim \mathcal{\overline{E}}(t_1) \quad\forall\quad I,\end{equation}
so that one may further approximate the spectral amplitude of the polarization to:
\begin{multline}\label{eq:pdump_a}
=\frac{\rho\pi\beta|\mathcal{E}(t_1)|^2}{8}\sum_{I,I',F}d_{0,I'}e^{ - i\frac{t_{I'}^2}{\beta}}d_{I,0}e^{ - i\frac{t_I^2}{\beta}}\\\overline{\mathcal{E}}(t_{F,I'})  e^{-i\frac{t_{F,I'}^{2}}{\beta}}\sqrt{\pi\beta}e^{i\frac{\pi}{4}}d_{I',F}d_{F,I}\\\frac{1}{E_F-E_I + \omega +\frac{i\Gamma_I}{2}} e^{-\frac{\Gamma_I (t_{F,I'} - t_I)}{2}}e^{-\frac{\Gamma_{I'} (t_{F,I'} - t_{I'})}{2}}
\end{multline}
Using the notation:
\begin{equation}
t_{I'} = t_{I} + \Delta
\end{equation}
\begin{equation}
t_{F,I'} = t_{F,I} + \Delta
\end{equation}
one may write the exponential factors:
\begin{equation}\label{eq:3expof}
e^{  i\frac{t_{I'}^2}{\beta}}e^{ - i\frac{t_{I}^2}{\beta}}e^{-i\frac{t_{F,I'}^{2}}{\beta}}
\end{equation}
as:
\begin{equation}
= e^{i\frac{(2\Delta t_{I} -2\Delta t_{F,I})}{\beta}}
e^{-i\frac{t_{F,I}^{2}}{\beta}}.
\end{equation}
Using \begin{equation}
\Delta = (E_I - E_{I'})\frac{\beta}{2} 
\end{equation}
one finds that
\begin{equation}
e^{  i\frac{t_{I'}^2}{\beta}}e^{ - i\frac{t_{I}^2}{\beta}}e^{-i\frac{t_{F,I'}^{2}}{\beta}}=
e^{-i(E_I-E_{I'})(t_{F,I} - t_{I})}
e^{-i\frac{t_{F,I}^{2}}{\beta}}
\end{equation}
so that 
\begin{multline}
\tilde{\mathcal{P}}_{Dump}(z,\omega) \sim\frac{\rho\pi\beta|\mathcal{E}(t_1)|^2}{8}\sum_{I,I',F}d_{0,I'}d_{I,0}\\e^{-i(E_I-E_{I'})(t_{F,I} - t_{I})}
\\\overline{\mathcal{E}}(t_{F,I'})  e^{-i\frac{t_{F,I}^{2}}{\beta}}\sqrt{\pi\beta}e^{i\frac{\pi}{4}}d_{I',F}d_{F,I}\\\frac{1}{E_F-E_I + \omega +\frac{i\Gamma_I}{2}} e^{-\frac{\Gamma_I (t_{F,I'} - t_I)}{2}}e^{-\frac{\Gamma_{I'} (t_{F,I'} - t_{I'})}{2}}
\end{multline}

Considering that the Fourier transform of the electric field can be approximated by:
\begin{multline}
\tilde{\mathcal{E}}(\omega) = \frac{1}{2}\int dt e^{i\omega t} \big[\mathcal{\overline{E}}(t)e^{-i(\omega_{0}t-\frac{t^{2}}{\beta})}+c.c.\big]\\\sim\frac{1}{2}\int dt e^{i\omega t} \big[\mathcal{\overline{E}}(t)e^{-i(\omega_{0}t-\frac{t^{2}}{\beta})}\big]\\
\sim \frac{1}{2}\mathcal{\overline{E}}(t_{\omega}) e^{-i\frac{t^2_{\omega}}{\beta}}\sqrt{\pi\beta}e^{i\frac{\pi}{4}}
\end{multline}
where $t_{\omega} = (\omega_0 - \omega)\frac{\beta}{2}$, one finds in the impulsive limit noting that for $\omega \sim E_I - E_F$, $t_{F,I}\sim t_{\omega}$:

\begin{multline}\label{eq:pdump_a}
\tilde{\mathcal{P}}_{Dump}(z,\omega > 0)\sim \frac{\rho\pi\beta|\mathcal{E}(t_1)|^2}{4}\sum_{I,I',F} \\ d_{0,I'}d_{I,0}e^{-i(E_I-E_{I'})(t_{F,I} - t_{I})}
\\\tilde{\mathcal{E}}(\omega) d_{I',F}d_{F,I}\\\frac{1}{E_F-E_I + \omega +\frac{i\Gamma_I}{2}} e^{-\frac{\Gamma_I (t_{F,I'} - t_I)}{2}}e^{-\frac{\Gamma_{I'} (t_{F,I'} - t_{I'})}{2}}
\end{multline}

In the impulsive limit, one may replace the excitation times $t_I$ by the average excitation time $t_1$ and the deexcitation times $t_{F,I}$ by the average excitation time $t_2 = \langle t_{F,I}\rangle_{F,I}$. 

This yields:
\begin{multline}\label{eq:pdump_a}
\tilde{\mathcal{P}}_{Dump}(z,\omega > 0)\sim \frac{\rho\pi\beta|\mathcal{E}(t_1)|^2}{4}\sum_{I,I',F} \\ d_{0,I'}d_{I,0}e^{-i(E_I-E_{I'})(t_{2} - t_{1})}
\\\tilde{\mathcal{E}}(\omega) d_{I',F}d_{F,I}\\\frac{1}{E_F-E_I + \omega +\frac{i\Gamma_I}{2}} e^{-\frac{\Gamma_I (t_{2} - t_1)}{2}}e^{-\frac{\Gamma_{I'} (t_{2} - t_{1})}{2}}
\end{multline}


\subsubsection{Spatial evolution of the spectrum}

We now insert the approximate expression for $\tilde{\mathcal{P}}_{Dump}(z,\omega > 0)$ given in Eq. \ref{eq:pdump_a}, in Eq. \ref{eq:mwe_SVEA2} to determine the spatial evolution of the spectrum:

 \begin{multline}\label{eq:mwe_SVEA3}
\frac{\partial\tilde{\mathcal{E}}(z,\omega)}{\partial z} \sim \frac{i\rho\pi^2\omega\beta|\mathcal{E}(t_1)|^2}{2c}\sum_{I,I',F} \\ d_{0,I'}d_{I,0}e^{-i(E_I-E_{I'})(t_{2} - t_{1})}
d_{I',F}d_{F,I}\\\frac{1}{E_F-E_I + \omega +\frac{i\Gamma_I}{2}} e^{-\frac{\Gamma_I (t_{2} - t_1)}{2}}e^{-\frac{\Gamma_{I'} (t_{2} - t_{1})}{2}}\times \tilde{\mathcal{E}}(z,\omega) \end{multline} 
This equation can be solved analytically yielding an exponential spatial dependence of the spectral amplitude $\tilde{\mathcal{E}}$. For the propagation-distance-dependent spectrum $|\tilde{\mathcal{E}}(z,\omega)|^2$, one obtains:
\begin{equation}
|\tilde{\mathcal{E}}(z,\omega)|^2 \sim |\tilde{\mathcal{E}}(z=0,\omega)|^2 e^{\sigma \rho z}
\end{equation}
where $\sigma$ denotes the emission cross section:
\begin{multline}
\sigma =  -\frac{\pi^2\omega\beta|\mathcal{E}(t_1)|^2}{c}\text{Im}\big[\sum_{I,I',F} \\ d_{0,I'}d_{I,0}e^{-i(E_I-E_{I'})(t_{2} - t_{1})}
d_{I',F}d_{F,I}\\\frac{1}{E_F-E_I + \omega +\frac{i\Gamma_I}{2}} e^{-\frac{\Gamma_I (t_{2} - t_1)}{2}}e^{-\frac{\Gamma_{I'} (t_{2} - t_{1})}{2}}\big]
\end{multline}


\bibliography{references}
\end{document}